\documentstyle[12pt]{article}
\begin{document}
%%%%preprint number%%%%%%
\begin{flushright} OCHA-PP-85 \\ October (1996) \end{flushright}
%%%%%%%%%%%%%%    
%%%%%%%%%%%%%%
\begin{center}
 {\Large\bf OLD-FASHIONED DUALITIES REVISITED\footnote{ Summary of lectures 
given at '96 Kashikojima Summer Institute on "Non-perturbative Aspects in 
Gauge Theories and the Phenomenological Applications" (August 27 $\sim$ 
September 1, 1996), Amagi-Highland Annual Seminars organized by Professor 
Seitaro Nakamura (July 5 $\sim$ 6), and the Yukawa Institute Workshop on 
"Quantum Field Theory - New Development of the Non-Perturbative Methods" 
(July 16 $\sim$ 19).}}\\
\bigskip
Akio SUGAMOTO\\
{\it Department of Physics, Ochanomizu University \\
2-1-1, Otsuka, Bunkyo-ku, Tokyo 112,  Japan}
 \end{center}    

\begin{abstract}
In the lectures given at '96 Kashikojima Summer Institute, Amagi-Highland 
Seminars, and Yukawa Institute Workshop, the old-fashioned dualities and 
the theory of extended objects are reviewed, starting from 't 
Hooft-Mandelstam duality and following the author's old works. 
Application of the old-fashioned dualities to new issues are also 
given on the D-branes, on the phase transition of B-F theory to 
Einstein gravity and on the swimming of microorganisms. 
\end{abstract}
%%%%%%%%%%%
\section{Introduction}
%%%%%%%%%%%
As you know in the Higgs mechanism, monopoles are united with the 
magnetic string, whereas in the Confinement mechanism, quarks are united with the electric string.  Both mechanisms are interchangeable by replacing the electric field with the magnetic field. This replacement is called "dual or duality transformation".

In the electrodynamics in the medium, we have the following Lagrangian
\begin{equation}
{ \cal L } =-\frac{1}{4}\varepsilon F_{\mu\nu}F^{\mu\nu}=\frac{1}{2}\varepsilon ({\bf E}^2 - {\bf B}^2 ) = \frac{1}{2} ({\bf E} \cdot {\bf D} - {\bf H} \cdot {\bf B} ) =-\frac{1}{2} \mu ({\bf H}^2 - {\bf D}^2 ) . 
\end{equation}
Here, $\varepsilon$ is the dielectric constant or the electric permeability.  Probably young people are more familiar with the notation $\tau$ rather than $\varepsilon$, after the work of Seiberg and Witten~\cite{S-W} has appeared.  
It is very important to note that the replacement of electric field with magnetic field induces in the medium, 
\begin{equation}
\varepsilon \leftrightarrow -\mu=- \frac{1}{\varepsilon } . 
\end{equation}
In the present fashion, this is the modular transformation $\tau \leftrightarrow  -\frac{1}{\tau} $.  It is the so called 't Hooft-Mandelstam Duality~\cite{'t Hooft Mandelstam} in which the completely symmetrical treatment  between Higgs and Confinement has been developed.

In late 70's, 't Hooft introduced in the  $SU(N_{c})$ gauge theories, $B(C', t) \equiv$ operator of creating a magnetic vortex ($\mbox{operator}^{\clubsuit}$ of creating a "bare soliton") along the curve $C'$ at time t .  It is the magnetic counterpart of the well-known Wilson operator $A(C, t) = Tr P \exp i\oint_{C} A_{\mu}(x)dx^{\mu} $ which produces an electric flux along the curve C.  

$\clubsuit$ A good reference at that time was the Mandelstam('75)'s work which constructs the operator of creating or annihilating the "bare soliton" in (1+1) dim. Sine-Gordon theory.
 \begin{equation}
{\cal L} = 1/2 \partial_{\mu }  \phi \partial^{\mu } \phi + \mu^{2}/\beta^{2} :\cos \beta \phi(x, t): .
\end{equation}
The operator $\hat{\psi } (x, t)$ of annihilating the bare soliton at position x at time t satisfies the following commutation relations:
\begin{equation}
[\hat{\phi}(x) , \hat{\psi}(y)] = \theta (x-y) 2\pi \beta^{-1} \hat{\psi}(x) , \enspace [\hat{\phi}(x) , \hat{\phi}(y)] = [\hat{\psi}(x) , \hat{\psi}(y)] = 0 . 
\end{equation}
Proof of these commutation relations is as follows: If the bare soliton (or anti-soliton) is defined to have the field configuration $\phi (y) = 2\pi \beta^{-1} n - (\mbox{or} +) \theta (x-y)$ with integer $n$, it is the solution having a positive (or negative) "gap" at position x .  Then, 
\begin{eqnarray}
\hat{\phi} (y) \hat{\psi} (x) | \phi (y) \rangle &=& \hat{\phi} (y) | \phi + \theta (x-y) 2 \pi \beta^{-1} \rangle \nonumber \\
&=& \phi (y) + \theta (x-y) 2 \pi \beta^{-1} | \phi + \theta (x-y) 2 \pi \beta^{-1} \rangle  \nonumber \\
&=& \hat{\psi} (x) \left( \hat{\phi} (y) + \theta (x-y) 2 \pi \beta^{-1} \right) | \phi (y) \rangle .
\end{eqnarray}
Here, two kinds of operators satisfying the above commutation relations are found:
 \begin{equation}
 \psi = \left( \begin{array}{c} \psi_1 \\
 \psi_2 \end{array} \right) \propto \left( \begin{array}{c} 
\exp -2\pi\beta^{-1} \int^{x}_{-\infty} dy \left( \dot{\phi} + \frac{\beta}{4\pi} \phi' \right) \\
 \exp -2\pi\beta^{-1} \int^{x}_{-\infty} dy \left( \dot{\phi} - \frac{\beta}{4\pi} \phi' \right) 
\end{array} \right) ,
\end{equation}
These are combined to form miraculously a fermion field in the massive Thirring model, 
\begin{equation}
{ \cal L } = \bar{\psi} ( -i\gamma^{\mu}\partial_{\mu} - m_0 )\psi + \frac{1}{2} g ( \bar{\psi} \gamma^{\mu}\psi)(\bar{\psi} \gamma_{\mu}\psi) ,
\end{equation}
where $g/\pi = 1- 4\pi/\beta^2 $.

The lesson obtained here: Bare soliton is given by the "singular" symmetry transformation of a system, which is $ \phi \rightarrow \phi +2\pi n \beta^{-1} ( n: {\rm integer}) $ in the above example.  The operator generating the bare soliton, introduces naturally a "gap" to the original field configurations.  Therefore, if such an operator has non-vanishing expectation values at some place, it means that the original variable changes very rapidly there.  The state having such rapidly changing configuration of the field is called "disordered state", in comparison with the ordered state where the original field variable has a constant expectation value.  The expectation value of the bare soliton creating operator is called "disorder parameter".  (In the above example, the disorder operator $\psi$ is a fermion, so that it cannot have a vacuum expectation value ... .)

Now, coming to the gauge theories, we perform "singular" gauge transformation (the "singular" symmetry transformation here), in the place where we want to produce a magnetic vortex.  Let's assume the circle $C$ (parametrized by $\theta$) winds around the vortex $C'$.  After travelling a turn ($\theta=0 \sim 2\pi$) around the vortex $C'$, the gauge transformation is assumed to change, according to
\begin{equation}
\Omega^{[C']} (\theta=2\pi) = \Omega^{[C']} (\theta=0) e^{2\pi i n/N} . 
\end{equation}
 The exponential factor here is an element of the center $Z_N$ of $SU(N_c)$.  The center constitutes all the elements which commute with all the $SU(N_c)$  elements, so that the elements of the center are proportional to the unit:
\newfont{\bg}{cmr10 scaled\magstep4}
\newcommand{\bigzerol}{\smash{\hbox{\bg 0}}}
\newcommand{\bigzerou}{%
\smash{\lower1.7ex\hbox{\bg 0}}}
\begin{equation}
Z_{N} = \left\{ \exp \left[ \frac{2\pi i n}{N} \left( \begin{array}{ccccc}
1&  &          &  &\bigzerou \\           &1&          & &    \\
  &  &\ddots& &     \\
  &  &          &1&    \\
  \bigzerol& & & &N-1 
\end{array} \right) \right] \right\}_{n=1 \sim N-1} . 
\end{equation}
Why is the element of the center adopted as the gauge transformation? The electric and magnetic fields which belong to the adjoint representation do not see the center, $\Omega(x)({\rm center}) F_{\mu\nu}(x) ({\rm center})^{-1}\Omega(x)^{-1} = \Omega(x) F_{\mu\nu}(x) \Omega(x)^{-1}$, so that magnetic vortex can not be generated by the gauge transformation, except at the singular place.  The gauge transformation is, however, singular along the curve $C'$, where the magnetic vortex ${\bf B}$ is produced.  The gauge transformation can be viewed Abelian-like, $\Omega^{[C']}(\theta) = e^{i \theta n/N}$ without the traceless generator, the induced gauge field reads $ \Delta A_{\mu}(x) = \frac{1}{g}\partial_{\mu}(\theta n/N)$, from which we find 
\begin{equation}
\int_{S} dS \enspace {\bf B} = \oint_{C} \Delta A_{\mu} dx^{\mu} = \frac{2\pi}{g} \frac{n}{N} .
\end{equation}
 Therefore, the singular gauge transformation $\Omega^{[C']}(\theta)$ generates the $Z_{N}$ vortex on the curve $C'$: 
\begin{equation}
\hat{B}(C', t)\mid A, \phi \rangle \equiv \mid A^{\Omega^{[C']}}, \phi^{\Omega^{[C']}}\rangle .
\end{equation}

Now we are able to prove the following't Hooft algebra: 
\begin{eqnarray}
& &\hat{A} (C) \hat{B} (C') \nonumber \\
&=&\hat{B} (C') \hat{A} (C) \times \cases{ e^{2\pi i n/N} & (\mbox{$C$ and $C'$ is entangled like a puzzle ring})\cr
 1  & (\mbox{otherwise}), \cr } \\
& &[\hat{A} (C), \hat{A} (C')] = [\hat{B} (C), \hat{B}( C')] = 0 .
\end{eqnarray}
Proof is the following:
\begin{eqnarray}
& &\hat{A}(C)\hat{B}(C')\mid A, \phi \rangle = \hat{A}(C) \mid A^{\Omega[C']}, \phi^{\Omega[C']} \rangle  \nonumber \\
&=& Tr P \exp i\oint_{C} A_{\mu}(x)^{\Omega[C']}dx^{\mu}\mid A^{\Omega[C']}, \phi^{\Omega[C']} \rangle  \nonumber \\
&=& Tr \{\Omega (2\pi) \left[ P \exp i\oint_{C} A_{\mu}(x)dx^{\mu} \right] \Omega(0)^{-1} \} \mid A^{\Omega[C']}, \phi^{\Omega[C']} \rangle \nonumber \\
&=& e^{2\pi i n/N} \hat{B}(C') \hat{A}(C) \mid A, \phi \rangle .
\end{eqnarray}
Of cource $[ \hat{A}(C), \hat{A}(C') ] = [ \hat{B}(C), \hat{B}(C') ] = 0 $ hold trivially.  
In the above commutation relations $A(C)$ and $B(C')$ appear completely symmetrically (like the mirror image as was stated by 't Hooft).  

Accordingly, if the Higgs phase exists:
\begin{equation}
 A(C) \sim \alpha_1 \exp (-\alpha_2 L(C))\enspace \mbox{and} \enspace B(C') \sim \beta_1 \exp (- \beta_2 \Sigma(C'), 
\end{equation}
then its mirror image, namely, Confinement phase exists as well: 
\begin{equation}
A(C) \sim \alpha_1 \exp (-\alpha_2 \Sigma(C))  \enspace \mbox{and} \enspace B(C') \sim \beta_1 \exp (- \beta_2 L(C').  
\end{equation}
Here, $L(C)$ and $\Sigma(C)$ represent the perimeter and the area depicted by the closed curve $C$, the cases of having $L(C)$ and $\Sigma(C)$, are called peimeter law and area law, respectively. Therefore, if the Higgs phase exists, so does the Confinement.

At the Kashikojima Summer Institute and afterwards, the author has learnt from Dr. M. Sakamoto and Dr. S. Nojiri the following important fact:

 Using the 't Hooft algebra, we can show that at least one of $A(C)$ and $B(C')$ (incliding the both) should obey the area law.  This is a very strong statement.  Dr. Sakamoto's question at Kashikojima was "What happens physically when both $A(C)$ and $B(C')$ give area laws?" The author's answer was "Dyon condensation probably occurs in that case."  We think it is a good lesson to study from the view point of 't Hooft algebra, what physically happens in the various phases, for example, in the tractable N=1 or 2 SUSY Yang-Mils theories.

%%%%%%%%%%%%%
\section{Dual Transformation}
%%%%%%%%%%%%%
%%%%%%%%%%%%%%
The author's impression on the 't Hooft algebra at that time ('78) was the following:
 
This is similar to the commutation relations in quantum mechanics:
\begin{equation}
[\hat{x},\hat{p}]=i , \enskip [\hat{x}, \hat{x}]=[\hat{p}, \hat{p}]=0 ,
\end{equation}
from which the uncertainty principle $(\Delta p)^2 \cdot (\Delta x)^2 \geq 1$ is derived.  In our case it can be a kind of uncertainty relationship that if the electric flux is squeezed, the magnetic flux is not, whereas the magnetic flux is squeezed, the electric fllux is not, namely, $(\Delta{\bf E})^2 \cdot (\Delta{\bf B})^2 \geq 1 $.  We know in quantum mechanics that the transformation from x-representation to p-representation (i.e. Fourier transformation) takes one theory to the other.  Therefore, the Fourier transformation with respect to the field strength is the dual transformation, taking one theory in the {\bf B}-representation to the other (dual) theory in the {\bf E}-representation.  In the course of the transformation, the interchange of $\varepsilon \leftrightarrow \mu $ is nicely incorporated.
Furthermore, we may naturally introduce the disorder operator such as $\psi(x)$ and $B(C,t)$.  

Then, we have considered the following transformation:
\begin{eqnarray}
&&\exp \left\{ i \int d^4x -\frac{1}{4} \varepsilon F^{a}_{\mu\nu}F^{a \mu\nu} \right\}  \nonumber  \hfill \\ 
&\propto&\int {\cal D}W^{a}_{\mu\nu}(x) \exp \left\{ i \int d^4x \left[ \frac{1}{4}m^2 \frac{1}{\varepsilon} W^{a}_{\mu\nu}W^{a\mu\nu}- \frac{1}{2}m \tilde{W}^{a}_{\mu\nu}F^{a\mu\nu} \right] \right\} ,
\end{eqnarray}
and the subsequent integration out of the original variable $A^{a}_{\mu}$.  (See references in~\cite{Abelian D-T}, ~\cite{non-Abelian D-T} and ~\cite{Savit}.) 
 Here a parameter $m$, having the mass dimension, may be absorbed into the definition of the anti-symmetric tensor field $W^{a}_{\mu\nu}(x)$.  It is noted that the above dual transformation can be carried out also in the non-Abelian models ~\cite{non-Abelian D-T}.  Here we will explain the dual transformation taking the simple Abelian Higgs model, following Ref.~\cite{Abelian D-T}.  

Original action reads
\begin{equation}
{\cal L} = -\frac{1}{4} F_{\mu\nu}F^{\mu\nu} + | (\partial_{\mu}+ i e A_{\mu}) \phi |^2 - V(|\phi|^2).
\end{equation}
With respect to the partition function of this model, the above dual transformation and the subsequent integration over $A_{\mu}$ are performed.  Then we have the action of the dual model:
\begin{eqnarray}
{\cal L^{\star}}&=& -\frac{m^2}{2 e^2 |\phi|^2} \frac{1}{2} (V_{\mu})^2 - \frac{1}{4} m^2 (W_{\mu\nu})^2 + (\partial_{\mu}|\phi|)^2 - V(|\phi|)  \nonumber \\
& & + \frac{1}{2} \frac{2 \pi}{e} m W^{\mu\nu} \times \frac{1}{4 \pi} \varepsilon_{\mu\nu\lambda\rho}(\partial^{\lambda} \partial^{\rho} - \partial^{\rho}\partial^{\lambda}) \chi(x) . 
\end{eqnarray}
Here, $V_{\mu} = \partial^{\nu} \tilde{W}_{\nu\mu}$ is the velocity field of the fluid, satisfying the contiuity equation, $\partial^{\mu}V_{\mu}=0$, and ${W}_{\mu\nu}$ is the velocity potential.  Let's call this model "Relativistic hydrodynamics of Kalb-Ramond-Nambu"~\cite{ K-R }.  Personally, the author did not understand at that time what is the physical meaning of the model so obtained in the above mentioned method.  Fortunately, he has a chance to listen the lecture given by Prof. Nambu at U. of Tokyo at Hongo campus (probably in the spring '78).  Then, he recognized that his result means the hydrodynemics.  
Recently this kind of model is rediscovered, and is applied to the study of vortices in the superfluidity~\cite{superfluidity}. 

Now, we will take into account the "singular" configuration, where the Higgs phase $\chi$ changes $2 \pi n$, after travelling a turn along $C$ winding $C'$.  The "singularity" seems to appear in the beginning (at the level of bare soliton or bare vortex), but it diappears after solving the equation of motion (at the level of real soliton or Nielsen-Olesen vortex).

In a simple case of $C'$ being on the $x^3$ axis, we have
\begin{equation}
\int dx^1 dx^2 ( \partial^1 \partial^2 -\partial^2 \partial^1 ) \chi = \int dS \enspace \mbox{rot} (\nabla \chi) = \oint_{C} \nabla_{\mu}\chi dx^{\mu}= \Delta \chi = 2\pi n, 
\end{equation}
that is,
\begin{equation}
\omega^{03} = \frac{1}{2\pi}(\partial^1 \partial^2 -\partial^2 \partial^1 ) \chi = n \delta(x^1) \delta(x^2) = n \int dy^0 dy^3 \delta^{(4)}(x-y) .
\end{equation} 
Therefore, if we parametrize the world sheet of the vortex generally as $y^{\mu}(\tau, \sigma)$, 
\begin{equation}
\omega^{\mu\nu}(x) = n \int d\tau d\sigma \frac{\partial(y^{\mu}, y^{\nu})}{\partial(\tau, \sigma)} \delta^{(4)}(x-y(\tau, \sigma)).
\end{equation}
From this, the coupling of the world sheet of the vortex with the Kalb-Ramond fields is derived, the contribution of which to the partition function reads
\begin{equation}
e^{i \int_S d\sigma^{\mu\nu} \frac{1}{2} \frac{2\pi}{e} m W_{\mu\nu}} .\label{KR interaction}
\end{equation}
The method stated above is the general way to introduce "bare soliton" and to obtain its coupling to the Kalb-Ramond field.  
Therefore, the method may be effective to introduce D-branes and to obtain their coupling to the Kalb-Ramond fields, since D-branes are the bare soliton of the string theory.  We will discuss this issue in the next section.  

Remember that the contribution to the partition function of the classical charged particle interacting with the electro-magnetic field is $e^{ i \int_C dx^{\mu} e A_{\mu}(x)}$.  
We know that if we sum up all the possible configurations with any number of closed paths depicted by the charged particles, we can incorporate any number of pair creations and subsequently occuring annihilations for the particles.  Then, the field theory of the charged particles, namely that of the charged Higgs model comes out. 
Similar thing happens if we perform "summing up all the possible shapes" of the world volume of the vortices. 
We want to incorporate the pair creation and the aubsequently occuring annihilation of the closed vortices, we sum up any number of the tori, depicted by the closed vortices.  

Therefore, we will consider the contribution to the partition function from the membrane (world volume of the closed vortex) which connects two closed paths, $C_1$ and $C_2$, with the area $A$.  Let $K( C_1, C_2; A )$ be the sum over "all possible shapes" of the membrane of the weight Eq. (\ref{KR interaction}), by fixing $C_1, C_2$, and $A$.  At that time ('80), we considered only the rectangular deformation of the menbrane with the side length $a$, and ignored the influence of the deformation of $C_2$ on $C_1$ itself and $C_2$.  This weak point should be remedied, by accomodating, for example, the works by Prof. H. Kawai and his collaborators on the random triangulation of 2 dim. surfaces with diffusion equation.  Then, the coupling of the vortex with the gravity may come in.  Our old-fashioned method may, however, be useful for getting the gross behavior of the result.    
Old-fashiondly, this $K$ is "proved" to satisfy the following diffusion equation:
\begin{equation}
\frac{\partial}{\partial \bar{A}} K( C_1, C_2; \bar{A}) = \hat{H}_{C_2} K( C_1, C_2; \bar{A}) ,
\end{equation}
where $a^{2} \bar{A}= A $ and $a$ is considered to a UV-cutoff.  Using the diffusion equation, the summing up the shapes of membranes gives the following dual action: 
\begin{eqnarray}
&{\cal L^{\star}}&=\sum_{C} \Psi[C]^{\dag} \hat{H}_{C} \Psi[C] \nonumber \hfill \\
&=&\sum_{C} \left\{ -\frac{1}{ \oint_{C}dx_{t}} \oint_{C} dx_t \biggl| \left( \frac{\delta}{\delta\sigma^{\mu t}} - i \frac{2\pi}{e} m W_{\mu t} \right) \Psi[C] \biggr|^{2} - M_{0}^{2} \biggl| \Psi[C] \biggr|^{2} \right\}.
\end{eqnarray}
 Here $M_{0}^{2}= [ 1-2(D-1)]/a^{4} < 0 $ (for the space-time dimension $D=4$) represents the entropy effect coming from the sum over the possible shapes ($D=4$), since there is no kinetic term of the vortex (or closed string) in our model. 

Notice: Role of the vacuum expectation value of the Higgs $\langle |\phi| \rangle^{2}$ is to give a mass to the gauge field $A_{\mu}$ in the original Higgs model, whereas it is to enhance the fluid velocity near the vortex in the dual hydrodynamics model: 
\begin{equation}
{\bf \nabla} \times (\frac{m^2}{2 e^2 \langle|\phi|\rangle^2}{\bf V} ) \propto {\bf \omega}_{\rm ext}.
\end{equation}

On the other hand, the vacuum expectation value of the string field $\langle \Psi[C] \rangle^2$ is to increase the mass of the Kalb-Ramond field in the dual hydrodynamics model, whereas it is to decrease the electric permeability $\varepsilon = (1+\psi_{S})^{-1}$, squeezing the electric flux connecting the external charges:
\begin{equation}
{\bf \nabla} \cdot (\varepsilon {\bf E}) \propto \rho_{\rm ext}.
\end{equation}
Here $\psi_{S}(x)$ represents the existence probability of the string passing through the point x , or the order parameter of the string condendsation:
\begin{equation}
\psi_{S}(x) \equiv 4 (\frac{2 \pi}{e})^2 \sum_{C(\ni x)}\langle |\Psi[C]|^2 \rangle / a^3 \oint_{C} dx_{t} .
\end{equation}
The effective potential for this $\psi_{S}(x)$ can be calculated in a similar way to the Coleman-Weinberg's work.  The result is given as follows:
\begin{eqnarray}
V(\psi_{S}) &=& \frac{1}{(4 \pi)^{2}} [(-20 e^2 \Lambda^4 + \frac{3}{2} \Lambda^2 m^2 ) \psi_{S}  \nonumber \hfill \\ 
& & -\frac{3}{4} m^4 \psi_{S}^2 \ln (\frac{\Lambda^2}{m^2}) + \frac{3}{4} m^4 (1+\psi_{S})^2 \ln (1+\psi_{S})] . \end {eqnarray} 

This method may be useful for obtaining the effective action for the D-branes.

Quark confinement in the strong coupling limit is shown by deriving the linear-rising potential between quark and anti-quark.  We don't understand why people talk about confinement without estimating the potential between quarks.  Therefore, we will show for the young people the old-fashioned way of calculating the quark potential.

When the external positive and negative charges are introduced, the Gauss law becomes
\begin{equation}
{\bf \nabla \cdot D} = [ \delta^{(3)} ({\bf x} +\frac{{\bf \ell}}{2}) - \delta^{(3)} ({\bf x} - \frac{{\bf \ell}}{2}) ] ,
\end{equation}
where ${\bf D}= \varepsilon {\bf E}$, and the electric permeability $\varepsilon$ becomes smaller due to the string condensation: $\varepsilon = (1+\psi_{S})^{-1}$.  This is connected to the dual Higgs mechanism (of generating the mass of the Kalb-Ramond field) by the Fourier transformation: 
\begin{equation}
-\frac{1}{4} \varepsilon F_{\mu\nu} F^{\mu\nu} \leftrightarrow -\frac{1}{4} \frac{1}{\varepsilon}W_{\mu\nu}W^{\mu\nu} .
\end{equation}
 Now the estimation of the potential between quark and anti-quark is to find the configuration of ${\bf D}$ and $\psi_{S}$, minimizing the following potential:
\begin{equation}
V_{Q \bar{Q}}(\ell) = \int d^3 x [ \frac{1}{2} (1+\psi_{S}) {\bf D}^2  + V(\psi_{S}) ] 
\end{equation}
There appears the $\psi_{S}^2$ term in the effective potential, so that we may write the potential for $\psi_{S}$ as 
\begin{equation}
V(\psi_{S}) = c_{e} \psi_{S} + d \psi_{S}^2 , 
\end{equation} 
where $c_{e} = -20 e^2 \Lambda^4 + \frac{3}{2}m^2$.  
In the strong coupling limit, the coefficient of the 1st order term in $\psi_{S}$ becomes negative and $\langle \psi_{S} \rangle \neq 0 $ (super state), but ${\bf D}^2$ also contributes to the same coefficient.  Therefore, if ${\bf D}^2$ becomes large, then $\langle \psi_{S} \rangle ¡á 0 $¡Ênormal state) is realized. 
This is the reason (dual Meissner effect) why the electric flux is squeezed inside the thin tube of the normal regin.  
Let the diameter and the length of the thin electric flux be $w$ and $\ell$, respectively, 
\begin{equation}
V_{Q \bar{Q}}(\ell) \sim \min_{w} \left[\pi w^2 \ell \left\{ \frac{1}{2} \left( \frac{Q}{\pi w^2 } \right)^2 + \frac{c_{e}}{4 d } \right\} \right] \sim \frac{Q^2}{\pi w_{\rm min}^2 }\times \ell.
\end{equation}
Under the above assumption, linear potential is obtained, but we should compare the result with that obtained under the other assumption of Coulomb-type potential.  After comparing both potentials, the smaller one is realized.  In our case, linear potential is realized when the distance between the quarks is less than the critical value, but if the distance exceeds the critical value, quarks may be liberated.  [See, Seo and Sugamoto in Ref.~\cite{Abelian D-T} for the details.]

%%%%%%%%%%%%%%%%
\section{Application to D-branes}
%%%%%%%%%%%%%%%%
%%%%%%%%%%%%%%%%
After reviewing the old-fashioned duality, we are probably tempting to apply it to the Kalb-Ramond field theory coupled with the string.  Recently, soliton solutions in the string theories are found~\cite{D-brane} and the duality in the string theories is actively studied using these soliton solutions~\cite{string duality}.

 Even high school kids know that there are two kinds of oscillation modes for the string, oscillation with free end point and that with fixed end point.  The boundary condition for the free end point is called Neumann B.C., and that of the fixed end point is called Dirichlet B.C..  The reason why the string theory with the fixed boundary condition has not been studied well is that we need a vessel like a barrel of the drum in order to fix the end points.  Recently we are able to incorporate the extended objects in the string theories, corresponding to the barrel of the drum, by calling "D-branes" borrowing "D" from the Dirichlet.

Let's consider an open string connecting two D-branes, where we assume that the both ends of the string can move inside the respective D-branes, but cannot leave from them.
If the both ends undergo a round trip inside the respective D-branes, the world sheet so obtained can be viewed in the other way as that one D-brane emits a closed string which is absorbed by the other D-brane afterwards.  Therefore, the D-brane is the source which can emit and absorb the closed string modes, and is the soliton solution of the string theory~\cite{D-brane}.  
This is just on the way from our old-fashioned duality in which bare soliton is constructed and the Kalb-Ramond theory coupled to the bare soliton is derived. 

Now, we will start with the Kalb-Ramond theory coupled to the string, which is the hydrodynamics model discussed in the previous section.  We take the stringy space-time dimension $D=10$.  Namely our starting action is 
\begin{equation}
S = \sum_{C} \sum_{x(\ni C)} \bigg|  \left( \frac{\delta}{\delta\sigma^{t\mu} (x) } - i A_{t\mu}(x) \right) \Psi[C] \bigg|^2 + \sum_{x} - \frac{\varepsilon}{2 \cdot 3! } F_{\mu\nu\lambda} F^{\mu\nu\lambda} - V(|\Psi[C]|^2) , 
\end{equation}
where $F_{\mu\nu\lambda}= \partial_{\mu}A_{\nu\lambda}+ \partial_{\nu}A_{\lambda\mu} + \partial_{\lambda}A_{\mu\nu}$
was rewritten as the velocity vector $V^{\mu}$ previously for $D=4$.  Here we consider the massless Kalb-Ramond field, so that the model has the Kalb-Ramond symmetry, the gauge symmetry of the string theory.
That is, 
\begin{eqnarray}
\Psi[C] &\rightarrow& \Psi'[C] = e^{i \oint_{C} dx^{t} A_{t}(x)} \Psi[C]  ,\nonumber \\
A_{\mu\nu}(x) &\rightarrow& A_{\mu\nu}'(x) = A_{\mu\nu}(x) + \partial_{\mu} \Lambda_{\nu}(x) - \partial_{\nu} \Lambda_{\mu}(x) .
\end{eqnarray}
Next we perform the old-fashioned dual transformation:
\begin{eqnarray}
&&\exp \left\{ i \int d^{10} x -\frac{\varepsilon}{2 \cdot 3!}F_{\mu\nu\lambda}F^{\mu\nu\lambda} \right\}   \nonumber \hfill \\
&\propto&\int {\cal D}W_{\mu_{1}\cdots \mu_{7}}(x) \hfill \\
&\exp&\left\{ i \int d^{10} x \left[ \frac{1}{2 \cdot 7!} \frac{1}{\varepsilon}  W_{\mu_{1} \cdots \mu_{7}} W^{\mu_{1} \cdots \mu_{7} } - \frac{1}{7! 3!} \epsilon^{\mu_{1} \cdots \mu_{10}} W_{\mu_{1} \cdots \mu_{7}} F^{\mu_{8} \mu_{9} \mu_{10}} \right] \right\}, \nonumber
\end{eqnarray}
and integrate out the original variables $A_{\mu\nu}(x)$ which was the gauge field $A_{\mu}$ previously.  
The string field $\Psi[C] = |\Psi[C]|  e^{ i \chi[C]}$ plays the same role as the Higgs field in the previous section.  Then, the dual action reads
\begin{eqnarray}
S^{\star} & = & \sum_{C} \left( \frac{\delta|\Psi[C]|}{\delta \sigma^{t\mu}(x)} \right)^2 + \sum_{C(\ni x)}|\Psi[C]|^2 \left( \frac{\delta\chi}{\delta\sigma^{t \mu}(x)} \right)^2 - \frac{\left( \sum_{C}|\Psi[C]|^2 \frac{\delta\chi}{\delta\sigma^{t \mu}(x)} \right)^2}{\sum_{C}|\Psi[C]|^2 } \nonumber \\
&  & - \frac{1}{8 \cdot 8! (7!)^2 } \frac{1}{\sum_{C(\ni x)} |\Psi[C]|^2}  F_{\mu_{1} \cdots \mu_{8}} F^{\mu_{1} \cdots \mu_{8}} \nonumber \\
&  & -\frac{1}{7! 2} \partial_{\nu} \left( \frac{-\frac{i}{2} \sum_{C(\ni x)} \Psi[C]^{\dag} \frac{\delta}{\delta\sigma^{\mu t}(x)} \Psi[C]}{\sum_{C(\ni x)}|\Psi[C]|^2 } \right) \epsilon^{12 \cdots \nu\mu t }\times W_{12\cdots 7} . 
\label{D-brane}
\end{eqnarray}
The last term is roughly
\begin{equation}
\sum_{x} \frac{1}{2} \sum_{C} \oint_{C} dx^{t} \frac{\delta^2 \chi [C]}{\delta\sigma^{\nu t}(x)\delta\sigma^{\mu t}(x)} \epsilon^{1, \cdots, 7\nu\mu t} \times \frac{1}{7!} W_{1, \cdots, 7}(x),
\end{equation}
which extracts the singularity of the
phase $\chi[C]$ in the string wave functional $\Psi[C]$ defind on the closed curve $C$.  

Let assume that the closed string $C$ evolves from a point, covers the surface of the 2 dim. sphere $S^2 $, and shrinks to the original point (Hopf fibration).  Accordingly the phase $\chi$ changes by $2 \pi n $ [n:integer].  Then, 
\begin{equation}
2 \pi n = \int\!\!\!\!\!\int_{S^2}\frac{\delta\chi[C]}{\delta\sigma^{\mu t}} dx^{\mu} dx^{t} .
\end{equation}
Here Stokes theorem (?) implys that there exists a singularity in the center of the $S^2 $. 
[Previously the torus $T^2$ formed by the transportation of the closed string $C$ around a point has been used, but the discussion of using $S^2 $ is more convenient for our purpose.]
Now, we have 
\begin{equation}
\oint_{C}dx^{t} \frac{\delta^2 \chi[C] }{\delta\sigma^{\mu t}(x) \delta\sigma^{\nu t}(x)} = 2 \pi n \delta (x^{\mu} - y^{\mu}) \delta (x^{\nu} - y^{\nu}) \delta (x^{t} - y^{t}) .
\end{equation}
The singularity is specified by the three delta functions in the 10 dimensional space-time, so that it forms a 7 dimensional surface.  
Namely, the last term of Eq.(\ref{D-brane}) is rewritten as 
\begin{equation}
\pi \sum_{\rm 7dim. world surface} \int \frac{1}{7!} d\sigma_{S_7}^{\mu_{1} \cdots \mu_{7}} W_{\mu_{1} \cdots \mu_{7}}(x) .
\end{equation}
This is a familiar coupling to the old-fashioned dualists.

Furthermore, performing the sum over "all the possible shapes" of the 7 dim. world surface where the phase of the string functional $\Psi [C]$ becomes singular, we can move from the original string theory coupled with the Kalb-Ramond field, to the following dual theory:  
\begin{eqnarray}
S^{\star} &= &\sum_{x} - \frac{1}{8 \cdot 8! (7!)^2 } \frac{1}{\sum_{C(\ni x)} |\Psi[C]|^2 } F_{\mu_{1} \cdots \mu_{7}} F^{\mu_{1} \cdots \mu_{7}} 
\nonumber \hfill \\
& &+\sum_{S_6} \sum_{x} \bigg| \left( \frac{\delta}{\delta\sigma^{\mu t_{1} \cdots t_{6}}(x)} - W_{\mu t_{1} \cdots t_{6}}(x) \right) \Psi[S_{6}] \bigg|^2
\nonumber\\
& &+\cdots . 
\end{eqnarray}
This is nothing but the Kalb-Ramond field theory where the 6-brane ($S_{6}$) interacts with its gauge field $W_{\mu_{1} \cdots \mu_{7}}$, and is a natural extension of the old-fashioned duality.
Similarly, we obtain the following correspondences between different brane theories, which are dually related with each other,
(where n-brane is the n-dimensionally extended object):
\begin{equation}
\begin{array}{cccc}
\mbox{0-brane} &\rightarrow & \mbox{7-brane} &(= \mbox{"vortex" connecting 6-branes "monopole"}) \\
\mbox{1-brane} &\rightarrow & \mbox{6-brane} &(= \mbox{"vortex" connecting 5-branes "monopole"})  \\
\mbox{2-brane} &\rightarrow & \mbox{5-brane} &(= \mbox{"vortex" connecting 4-branes "monopole"})  \\
\mbox{3-brane} &\rightarrow & \mbox{4-brane} &(= \mbox{"vortex" connecting 3-branes "monopole"})
\end{array}
\end{equation}
The method stated above has derived, by using the old-fashioned dual transformation, the dual theory of the "vortex-brane" connecting connects the "monopole-branes". 
Therefore, to obtain the dual theory with "monopole -branes", we have to move to the one dimensionally lower brane theory. 
This reasoning of the discrepancy by one dimension in the dual correspondence between ours and others, is due to the discussion with Dr. N. Ishibashi at the Chubu Summer Institute at the end of this August. The details of this section will be clarified in~\cite{Seo Sugamoto}.

%%%%%%%%%%%%%%%%%%
\section{Non-Abelian Dual Transformation}
%%%%%%%%%%%%%%%%
%%%%%%%%%%%%%%%%%
As was stated in the beginning, the dual transformation can be carried out also in the non-Abelian gauge theories.  Examples in the $SU(2)$ gauge theories are found in 

1) the model with two Higgs triplets, having the vortex solution, (Seo, Okawa and Sugamoto ('79) in Ref.~\cite{non-Abelian D-T}),

2) the model with one Higgs triplet, having the monopole solution, and the $SU(2)$ pure Yang-Mills model. (Seo and Okawa ('80) in Ref.~\cite{non-Abelian D-T})

Especially, dual transformation of $SU(2)$ pure Yang-Mills theory by Seo and Okawa~\cite{non-Abelian D-T} is very beautiful.

In these models, the dual models are Freedman's non-Abelian generalization of the Kalb-Ramond-Nambu theory:
\begin{equation}
{\cal L}^{\star} = -\frac{1}{2}\left( \frac{m}{e} \right)^2 \left( V^{a,\mu} - \frac{e}{m} j^{a,\mu} \right) M^{ab}_{\mu\nu} \left(V^{b,\nu} - \frac{e}{m} j^{b,\nu} \right) - \frac{1}{4} m^2 (W^{a}_{\mu\nu})^2 + \cdots .  \label{non-Abelian D-T}
\end{equation}
Here, $V^{a}_{\mu} = \partial^{\mu} \tilde{W}^{a}_{\nu\mu}$ is the velocity vector of the non-Abelian fluid, the current $j^{a,\mu}$ is given model by model, and $M^{ab}_{\mu\nu} \equiv (K^{ab}_{\mu\nu})^{-1}$ is the inverse matrix of $K^{ab}_{\mu\nu}$ which is the coefficient of the 2nd order terms in the gauge fields:
\begin{equation}
K^{ab}_{\mu\nu} = \frac{1}{2}   \sum_{\rm i-th Higgs}(\phi^{\dag}_{i} T^{\{ a}_{i}  T^{b \}}_{i} \phi_{i}) g_{\mu\nu} - \frac{m}{e}\epsilon^{abc} \tilde{W}^{c}_{\mu\nu} .
\end{equation}
The first term in the above action (\ref{non-Abelian D-T}) is invariant under the non-Abelian Kalb-Ramond transformation, 
\begin{equation}
W^{a}_{\mu\nu} \rightarrow W^{a}_{\mu\nu} + \nabla^{ab}_{\mu}\Lambda^{b}_{\nu} - \nabla^{ab}_{\nu}\Lambda^{b}_{\mu},
\enspace \nabla^{ab}_{\mu} = \delta^{ab} \partial_{\mu} + \frac{m}{e} \epsilon^{acb} P^{c}_{\mu}. \label{K-R symmetry} 
\end{equation}
The gauge field $P^{a}_{\mu}(x)$ appeared in the covariant derivative of the Kalb-Ramond transformation, is the dual gauge field having a complex expression: $\hat{P}^{a}_{\mu}(x) = M^{ab}_{\mu\nu}(\phi, W) V^{d,\nu}$, but the equation of motion shows a natural dual gauge field structure:
\begin{equation}
-(\frac{m}{e})^2 [\partial_{\mu} \hat{P}^{a}_{\nu} - \partial_{\nu} \hat{P}^{a}_{\mu} + \frac{m}{e} \epsilon^{abc} \hat{P}^{b}_{\mu} \hat{P}^{c}_{\nu}] = m^2 \tilde{W}^{a}_{\mu\nu} .
\end{equation}

 Furthermore, the work by Freedman and Townsend~\cite{Freedman-Townsend} has appeared, in which the Kalb-Ramond theory is related to the non-linear sigma model.
For example, starting from the following action,
\begin{equation}
{\cal L} = - \frac{1}{2} m^2 (A^{a}_{\mu})^2 - 2 m \tilde{W}^{a}_{\mu\nu} F^{a}_{\mu\nu}, 
\end{equation}
if we solve the equation of motion for $W^{a}_{\mu\nu}$, we obtain $F^{a}_{\mu\nu} = 0$.  Substituting the solution $ t^{a}A^{a}_{\mu} = \frac{1}{i g} U \partial_{\mu} U^{-1}$ 
into the original action gives the non-linear sigma model:
\begin{equation}
{\cal L} = - \frac{m^2 }{g^2 } Tr ( \partial_{\mu} U^{-1} \partial^{\mu} U ).
\end{equation}
They have studied the supersymmetric non-linear sigma model as well.  Details can be found in the original paper.
%%%%%%%%%%%%%%%%%%%
%%%%%%%%%%%%%%%%%%%
\section{Membrane or (n-1)-Dimensionally Extended Objects}
%%%%%%%%%%%%%%%%%%%
%%%%%%%%%%%%%%%%%%%
We once called the present p-brane, the (n-1)-dimensionally extended objects, with (n-1) = p.
As the action for such (n-1)-dimensionally extended objects, we have considered firstly 

(1) Nambu-type action:
\begin{equation}
S_{(1)} \propto \mbox{world volume} \propto \int d^{n}\xi \sqrt{ \frac{1}{n!} \left( \frac{\partial (x_{\mu_{1}} \cdots x_{\mu_{n}}) }{\partial (\xi_{1} \cdots \xi_{n})} \right)^2 }, 
\end{equation}
where $( \xi_{1} \cdots \xi_{n})$ is the parametrization of the world volume.  

Following Nambu, we will write the Jacobian as  
\begin{equation}
\{ x_{\mu_{1}}, \cdots, x_{\mu_{n}} \} \equiv \frac{\partial (x_{\mu_{1}} \cdots x_{\mu_{n}}) }{\partial (\xi_{1} \cdots \xi_{n})} \equiv p_{\mu_{1}, \cdots, \mu_{n}}. 
\end{equation}
We have considered secondly 

(2) Gravity-like action:
\begin{equation}
S_{(2)} \propto \int d^n \xi \sqrt{g(x)} \{(1-\frac{n}{2}) + \frac{1}{2} g^{ab} \partial_{a} x^{\mu} (\xi) \partial_{b} x_{\mu} (\xi) \}. 
\end{equation}
(See, Sugamoto ('83)~\cite{membrane}.)
Personally, the author has named his paper "Theory of Membranes", since the membranes has been mainly studied, but it has been the "Theory of p-Branes" in the present terminology.  Though he was interested in the generalization of the string theories, he was tempted to construct the theory of biological cells, being able to predict the probability for the fission and fusion of them.  It is regretable that since 15 years have passed, he still has to leave the issue in the future study.  In the paper, he has written a few other models for the membranes, among which is 

(3)  the action using the extra parameters:

Based on the Nambu's idea~\cite{Nambu}, we can paramerize the whole $D$ dim. space-time as $( x_{1}, \cdots, x_{D}) \equiv (\xi_{1}, \cdots, \xi_{n}; \phi_{1}, \cdots, \phi_{D-n})$, by adding extra parameters $(\phi_{1}, \cdots, \phi_{D-n})$ to the parametrization $( \xi_{1} \cdots \xi_{n})$ of the (n-1)-dimensionally extended object.  

Then, the third action ${\cal L}^{\star} ( \phi_{1}(x), \cdots, \phi_{D-n}(x))$ is so derived as that the n-dimensional manifold, obtained by setting $\phi_{1}(x), \cdots, \phi_{D-n}(x) = \mbox{constants} $ in the equation of motion of the action, coincides with the world volume determined by the dynamics of the (n-1)-dimensionally extended object.  
In the case of $D=4$ and $n=3$ (or $p=2$), it is enough to introduce one additional scalar, and the third action reads 
\begin{equation}
{\cal L} \propto \sqrt{(\partial_{\mu}\phi)^{2}}.
\end{equation}
This action in~\cite{membrane} has been rediscovered recently by Dr. Jens Hoppe~\cite{Hoppe}.
To derive this type of action the  following identity has been used:
\begin{equation}
p_{\mu_{1}, \cdots, \mu_{n}} \equiv \frac{\partial (x_{1}, \cdots, x_{D})}{\partial (\xi_{1}, \cdots, \xi_{n}, \phi_{1}, \cdots, \phi_{D-n})} \frac{\partial(\phi_{1}, \cdots, \phi_{D-n})}{\partial(x_{\nu_{1}}, \cdots, x_{\nu_{D-n}})} \times \varepsilon_{\mu_{1}, \cdots, \mu_{n}, \nu_{1}, \cdots, \nu_{D-n}}.
\end{equation}
The fourth action has been written in 

(4) the Hamilton-Jacobi formalism:

This also starts from Nambu's idea~\cite{Hamilton-Jacobi}.  Applying his idea to membrane, variation of the "action" should be the 3-form:
\begin{equation}
\sum_{m=1, 2} dS_{m} \wedge dT_{m} \wedge dU_{m} = \sum p_{\mu\nu\lambda} dx^{\mu}\wedge dx^{\nu} \wedge dx^{\lambda} - H d\xi_{1} \wedge d\xi_{2} \wedge d\xi_{3} .
\end{equation}
This is a natural extension of the usual $dS = p dx - H dt $ for the point particle.  The equation of motion in this formalism reads
\begin{eqnarray}
\{x^{\mu}, x^{\nu}, x^{\lambda}\} & = & \frac{\partial H}{\partial p_{\mu\nu\lambda} }, \\
\sum_{\nu > \lambda} \{ p_{\mu\nu\lambda}, x^{\nu}, x^{\lambda} \} & = &- \frac{\partial H}{\partial x^{\mu}} .
\end{eqnarray}
Here $d\xi_{1} \wedge d\xi_{2} \wedge d\xi_{3}$ is the "time" evolution parameter.  It is $d\xi_{1} \wedge d\xi_{2}
= d\tau \wedge d\sigma$ in the string theory.  
In this view, the modular transformation: $\tau \rightarrow -\frac{1}{\tau}$ or T-duality: $\tau \leftrightarrow \sigma$ can be naturally understood as the discrete transformation of $\xi_{1}$ and $\xi_{2}$, keeping $d\xi_{1} \wedge d\xi_{2}$ to be invariant.
It is an interesting problem to study what is T-duality in the membrane or p-branes.  Personally, one of the problems given in the author's lecture~\cite{Summer School} on the heterotic string theory at the 33th Summer School for Youth ('87), was the following:

 [Problem 5] "Can you generalize the modular invariance-like concept to the loop amplitude for membrane and the more extended objects ?" 

In considering these problems, it is interesting to study the discrete transformation of $(\xi_{1}, \cdots, \xi_{p+1})$ keeping $d\xi_{1} \wedge \cdots \wedge d\xi_{p+1}$ to be invariant in the Nambu's Hamilton-Jacobi formalism.

%%%%%%%%%%%%%%%%%%%%
\section{Gravity and Dual Meissner Effect}
%%%%%%%%%%%%%%%%%%%
%%%%%%%%%%%%%%%%%%%
As we know that the Einstein gravity looks very similar to the gauge theory, but an essential difference exists.
The curvature tensor $R^{\mu}{ }_{\nu, \lambda\rho}$ is written in terms of the Christoffel's symbol $\Gamma^{\mu}{ }_{\nu, \rho}$:
\begin{equation}
R^{\mu}{ }_{\nu, \lambda\rho} = \partial_{\lambda} \Gamma^{\mu}{ }_{\nu, \rho} - \partial_{\rho} \Gamma^{\mu}{ }_{\nu, \lambda}+ \Gamma^{\mu}{ }_{\sigma, \lambda}  \Gamma^{\sigma}{ }_{\nu, \rho} - \Gamma^{\mu}{ }_{\sigma, \rho}  \Gamma^{\sigma}{ }_{\nu, \lambda}, 
\end{equation}
where $\Gamma^{\mu}{ }_{\nu, \rho}$ is the gauge field, having two types of induces; $SO(1, 3)$ indices $\mu$ and $\nu$ (or $SO(4)$ indices in the following discussion) which may be called "isospace" indices, and the other type of the "real space" index $\lambda$.  After expressing the Christoffel's symbol in terms of the metric:
\begin{equation}
\Gamma^{\mu}{ }_{\nu, \rho} = \frac{1}{2} g^{\mu\mu'} (-\partial_{\mu'} g_{\nu\rho} + \partial_{\nu} g_{\mu'\rho} + \partial_{\rho} g_{\mu'\nu}), 
\end{equation}
the "isospace" and the "real space" indices are mixed.  The situation resembles that of monopole in the $SO(3)$ Higgs model, where the isospin index is connected to the space-time vector index in the hedgehog-like way.  In the present case, the metric $g_{\mu\nu}(x)$, or the vierbein $e^{A}_{\mu}(x)$ plays the role of a messenger connecting the isospace and the real space.  

Therefore, it is interesting to consider that the isospace and the real space are originally different concepts, but later they are mixed somehow, hopefully by the Higgs-like mechanism.  

The derivation of the Einstein gravity dynamically from the theory without a metric has a long history as the  "pregeometry".  Please refer to Ref.~\cite{Terazawa} for this approach.  We will consider in the following, the condensed medium of the strings or polymers from which the metric will come out, so that our approach might somehow be related to the "metrical elasticity" of A. D. Sakharov. (This was pointed out by Prof. K. Akama several years ago when a preliminary version of this section was presented informally at INS.)  

Now, we start from the topological 2-form gravity given by the action~\cite{B-F}
\begin{equation}
S=\int d^{4}x \frac{1}{2} \epsilon^{\mu\nu\lambda\rho} B_{\mu\nu}^{AB}(x)R_{\lambda\rho}^{AB}(x),      \label{B-F}
\end{equation}
and derive the Einstein action through the condensation of the strings (the dual Higgs mechanism, or the dual Meissner effect)~\cite{condensation of string}.
The theory given by Eq. (\ref{B-F}) is sometimes called "B-F theory", the name of which comes from the couplig of the anti-symmetric tensor field $B^{AB}_{\mu\nu}$ with the Field strengh,  $R_{\lambda\rho}^{AB}(x) = \partial_{\lambda}\omega_{\rho}^{AB}(x)-\partial_{\rho}\omega_{\lambda}^{AB}(x)+\omega_{\lambda}^{AB}(x)\omega_{\rho}^{CB}(x)-\omega_{\rho}^{AC}(x)\omega_{\lambda}^{CB}(x)$, given in terms of the $SO(4)$ gauge field called the spin connection $\omega_{\mu}^{AB}(x)$.
In our old-fashioned duality, the action is nothing but the Fourier transform factor of the non-Abelian dual transformation, that is,
\begin{equation}
S = \int d^{4}x \tilde{W}^{AB}_{\mu\nu} F^{AB\mu\nu}.
\end{equation} 
Therfore, it is invariant under the $SO(4)$ version of the non-Abelian Kalb-Ramond symmetry in Eq. (\ref{K-R symmetry}):
\begin{equation}
B_{\mu\nu}^{AB} \rightarrow B_{\mu\nu}^{AB} +\nabla_{[\mu}^{AC}(\omega)\Lambda_{\nu]}^{BC}(x) .  
\label{K-R}
\end{equation}

Let's compare "B-F" theory with the Einstein gravity.

A) B-F theory:
1) without metric, 2) larger symetry, 3) topological (non-dynamical);

B) Einstein gravity:
1') with metric, 2') smaller symmetry, 3') dynamical.

Extra Kalb-Ramond symmetry in the  B-F theory is the origin of the larger symmetry than the Einstein theory.
If we set 
\begin{equation}
B_{\mu\nu}^{AB}(x) = \frac{1}{2} \epsilon^{ABCD}e_{\mu}^{C}e_{\nu}^{D},  \label{the special form}
\end{equation}
by choosing a special form for $B_{\mu\nu}^{AB}$ written in terms of $e_{\mu}^{C}$, and by discarding 20 degrees of freedom, we arrive at the Einstein gravity.  Accordingly, the Kalb-Ramond symmetry is broken.  Topological nature of the B-F theory can be understood, if we rewrite the action in Eq. (\ref{B-F}) as follows:
\begin{equation}
S = \int \frac{1}{2} \epsilon^{ijk} \left( B^{a}_{0i} R^{a}_{jk} + B^{a}_{jk} R^{a}_{0i} \right),
\end{equation}
then $\omega^{a}_{i}$ and $B^{a}_{jk}$ are dynamical variables conjugate with each other, but $B^{a}_{0j}$ and $\omega^{a}_{0}$ are non-dynamical.
The equations of motion for the non-dynamical variables, give the constraints on the wave functional:
\begin{equation}
R^{a}_{jk} \Psi[\omega^{a}_{i}] = 0,\enspace \mbox{and}\enspace \nabla_{i} B^{a}_{jk} \Psi[\omega^{a}_{i}] = 0. 
\end{equation}
From the first constraint the gauge field becomes the pure gauge field, having no physical degrees of freedom.  The constraints indicate also the Kalb-Ramond invariance and the $SO(4)$ gauge invariance.  Especially, $R^{a}_{jk}$ can be viewed as the generator of the Kalb-Ramond transformation, namely, 
\begin{equation}
  R^{a}_{jk} = \nabla_{j}(\omega) \frac{1}{2} \epsilon_{klm} \frac{\delta}{\delta B^{a}_{lm}} - \nabla_{k}(\omega) \frac{1}{2} \epsilon_{jlm} \frac{\delta}{\delta B^{a}_{lm}} .
\end{equation} 
Therefore, owing to the extra Kalb-Ramond symmetry, all the physical degrees of freedom disappear, or the theory becomes a topological one. (For more details, see Ref.~\cite{proof of topological}.) Then, Dr. H. Nakano asked at Kashikojima what the observables of the B-F thoery are.  The author replied that they may be the linking number of the closed curve $C$ and of the closed surface $S$ (See, Ref.~\cite{non-Abelian linking}).  
 
Let's decompose $SO(4)$ with index $(AB)$, into $SU(2)$ with index $a$ times $SU(2)'$ with index $a'$, that is, \begin{equation}
a = (a4) + (bc), \enspace \mbox{and} \enspace a' = (a4) - (bc),  \label{self-dual decomposition}
\end{equation}
where $(abc)$ is the cyclic permutation of $(123)$, and of course $a, a'= 1 \sim 3$.  Then, the 20 constraints to guarantee the special form Eq.(\ref{the special form}), are given by
\begin{eqnarray}
\epsilon^{\mu\nu\lambda\rho}[B_{\mu\nu}^{a}B_{\lambda\rho}^{b}- \frac{1}{3}\delta^{ab}B_{\mu\nu}^{c}B_{\lambda\rho}^{c}] & = & 0,
\label{C1}\\
\epsilon^{\mu\nu\lambda\rho}[\bar{B}_{\mu\nu}^{a'}\bar{B}_{\lambda\rho}^{b'}-\frac{1}{3}\delta^{ab}\bar{B}_{\mu\nu}^{c'}\bar{B}_{\lambda\rho}^{c'}] & = & 0,
\label{C2}\\
\epsilon^{\mu\nu\lambda\rho}B_{\mu\nu}^{a}\bar{B}_{\lambda\rho}^{b'} & = & 0 , 
\label{C3} \\
\epsilon^{\mu\nu\lambda\rho} [B_{\mu\nu}^{a}B_{\lambda\rho}^{a} + \bar{B}_{\mu\nu}^{a'}\bar{B}_{\lambda\rho}^{a'}] & = & 0.
\label{C4}
\end{eqnarray}
If we consider the first $SU(2)$ part by setting $\bar{B}^{a'}_{\mu\nu} = 0$, then the self-dual part remains, that is, the $SO(4)$ indices satisfy $(a4)=(bc)$ with $(abc)$ the cyclic permutation of $(123)$.  For this self-dual part, the constraint becomes Eq.(\ref{C1}), namely
\begin{equation}
\mbox{traceless part of}  [\epsilon^{\mu\nu\lambda\rho}B_{\mu\nu}^{a}B_{\lambda\rho}^{b}] = 0,
\label{chiral constraint}
\end{equation}
Therefore, including this constraint Eq.(\ref{chiral constraint}) using the Lagrange multiplier field $\phi^{ab}(x)$, the action becomes,
\begin{equation}
  S = \int \frac{1}{2} \epsilon^{\mu\nu\lambda\rho}
\left( B^a_{\mu\nu}(x)R^a_{\lambda\rho}(x) + 
\underbrace{ \phi^{ab}(x)B^a_{\mu\nu}(x)B^b_{\lambda\rho}(x)}_
{constraint\: term} \right).
\label{chiral action}
\end{equation} 
This is called the "2-form gravity" which becomes the Einstein gravity if the constraint Eq.(\ref{chiral constraint}) as well as the equation of motion for the spin connection are solved in terms of the vierbeins~\cite{ 2-fom gravity }.

Since the Kalb-Ramond symmetry is the gauge symetry of the string, we can introduce the string field $\Psi[C; x_0]$, in the Kalb-Ramond invariant manner and discuss its spontaneouly breakdown through the string condensation.  

Our starting action is 
\begin{eqnarray}
  S&=& \int d^4 x\frac{1}{2} \epsilon^{\mu\nu\lambda\rho}
B^a_{\mu\nu}(x)R^a_{\lambda\rho}(x) \nonumber \\ 
& & + \sum_C \sum_{x_0 (\in C)} \sum_{x( \in C )} \epsilon^{\mu\nu\lambda\rho}
\left[ \left( \frac{\delta}{\delta C^{\mu\nu}(x)} + 
T^a B^a_{\mu\nu}[C; x, x_0]\right) \Psi[C; x_0] \right]
 ^{\dagger}\nonumber \\
& &~~~\times \left[  \left( \frac{\delta}{\delta C^{\lambda\rho}(x)} + 
T^a B^a_{\lambda\rho}[C; x, x_0]\right) \Psi[C; x_0]  \right] \nonumber \\
& &+ \sum_{C}\sum_{ x_0}  V [ \Psi[C; x_0] ^{\dagger} \Psi[C; x_0]] .
\label{ non-Abellian K-R action } 
\end{eqnarray}
In this expression we need to modify the Kalb-Ramond field $B^a_{\mu\nu}$ and its transformation to the non-local ones, which relfects the complexity of the non-Abelian Kalb-Ramond theory: 
\begin{eqnarray}
\Psi^i[C;x_0] & \rightarrow & 
U[C;x_0]^i_j\Psi^j[C;x_0],\\ \label{non-Abelian K-R 1}
B_{\mu\nu}[C; x, x_0] & \rightarrow &  U[C;x_0] B_{\mu\nu}[C; x, x_0] U[C;x_0]^{-1} - \frac{\delta U[C;x_0]}{\delta C^{\mu\nu}} U[C;x_0]^{-1},
\label{non-Abelian K-R 2}
\end{eqnarray}
where the non-Abelian transformation matrix depending on the curve $C$ reads:
\begin{equation}
U[C;x_0] \equiv  \left[ \exp \left\{ 
 i \oint_C dx^{\mu} \Lambda^{a}_{\mu}(x) W[ x\leftarrow x_0]^a_b
 T^b_R \right\} \right]^i_j .
\label{non-Abelian K-R 3}
\end{equation}
Here, Wilson operator on the path $P$ along the curve $C$,  
\begin{equation}
   W_P [x\leftarrow x_0]^a_b = \left[ P \: \exp i 
\int_{P( x\leftarrow x_0)} dx^\mu ( \Lambda_\mu^c T_A^c ) 
\right]^a_b 
 \label{ Wilson operator}  
\end{equation}
should be introduced (see Ref.\cite{condensation of string} for the details).

 Now the condensation of the string fields 
\begin{equation}
 \frac{1}{2}\phi^{ab}[C; x_0] \equiv \langle\Psi[C; x_0] ^{\dagger} T^a T^b\Psi[C; x_0] \rangle,
\label{non-Abelian condensation}
\end{equation}
plays the role of the Lagrange multiplier.  If the condensation becomes large for the symmetric (isospin 2) part of $(a, b)$, or the dual Meissner effect undertakes, then the constraint in Eq.[\ref{chiral constraint}] appears,  leading to the Einstein gravity. Otherwise, the theory stays in the topological phase.

The mechanism of the string field condensation stated above is very similar to the quark confinement mechanism discussed in Section 2 from the view point of the dual transformation.  Both mechanisms of the generation of metric and of the quark confinement are formally interchanged, by the replacement of the roles of the epsilon symbol $\epsilon^{\mu\nu\lambda\rho}$ and the product of the metric, $g^{\mu[\lambda}g^{\nu\rho]}$.  But, the condensation of the string field may commonly underlie the both phenomena.  
 In the polymer physics (a kind of string theory with some interactions), a similar condensation mechanism of the spin 2 field exists~\cite{polymer}.  If we define ${\bf u}$ as a unit vector representing the direction of the monomer, then the expectation value $\langle (u^a u^b) \rangle$ ($a = 1\sim 3$) is non-vanishing when the stress, given by the stress tensor $\sigma^{ab}$, is applied from outside to the polymer solution: We have
\begin{eqnarray}
   Q^{ab} &\equiv& \langle (u^a u^b - \frac{1}{3} \delta^{ab})\rangle \\
&\propto&  (\sigma^{ab} - \frac{1}{3} \delta^{ab}\sigma^c_c)\\
&\propto&  (\varepsilon^{ab} - \frac{1}{3} \delta^{ab}\varepsilon^c_c),
\label{polymer}
\end{eqnarray}
where $Q^{ab}$ is called "directional order parameter" and $\varepsilon^{ab}$ is the dielectric permeability of the polymer solution.  The last equation gives the optoelasticity, which can be undersood as the generation of the metric or of the distortion of the space, causing the double refraction, or the gravitational lensing effect in our terminology.  The dynamical origin of such an interesting phenomena is "the entropy effects" ( the effect of  summing up all the possible shapes of polymers) as well as the interactions between the monomers such as the nematic interaction of the liquid crystals.
Therefore, in order to understand the gravity, especially the generation of the metric, we should persue the condensation mechanism of the extended objects (the string fields) based on the entropy effects as well as the interactions between the portions of the strings (the monomers), following the polymer dynamics.

%%%%%%%%%%%%%%%%%
%%%%%%%%%%%%%%%%%%
\section{Duality between Microorganism's Swimming and String -Membrane Theory}
%%%%%%%%%%%%%%%%%%
The relativistic hydrodynamics of Kalb, Ramond and Nambu is a model of hydrodynamics coupled to the string and membrane variables.  Recently, we have studied the swimming of microorganisms in the terminology of string and membrane theories, by viewing that the surface of the microorganism $X^{\mu}(t; \xi_{1},\cdots, \xi_{n})$ is considered as a variable of closed string ($n=1$ ; 2 spacial dimensions¡Ëor membrane ($n=2$ ; 3 spacial dimensions) which is coupled to the velocity field of the external fluid~\cite{ours}. 

There is an amazing scaling law~\cite{Motokawa} between the swimming velocity $V$ and the length $L$ of the swimmer:
\begin{equation}
V = \left\{ 
\begin{array}{ll}
L / [s], &\mbox{(for swimmer in the water),} \nonumber \\
20 L / [s], &\mbox{(for swimmer in the air).}
\end{array}
\right.
\label{scaling law}
\end{equation}

This is valid for the organic as well as the inorganic swimmers; from microorganisms to submarines in the water, and from flies to aircrafts in the air.  We think that clarification of such scaling law should be another theme of our physicists.  
The above mentioned scaling law tells us that the Reynolds number $R$, giving the ratio of the kinetic term over the viscosity term in the equation of motion, takes a simpler form,
\begin{equation}
R = (L / [mm])^2,
\end{equation}
for both swimmers in the water and in the air.  Here $R=\rho V L / \mu$, the density $\rho$ is $1 (10^{-3}) [g/cm^3]$ for the water (air), and the viscosity coefficient $\mu$ is  $10^{-3} (20 \times 10^{-6}) [Pa \cdot s]$ for the water (air).

Now, the swimming motion in the water of the microorganisms, having the lengh much smaller than 1 [mm], is  a very special one.  Since the Reynolds number $R \ll 1$, the microorganisms feel the water as a very sticky or elastic material, something like the strings or the membranes.  Then, their equations of motion read  
\begin{equation}
{\bf \nabla}{\bf v}=0, \enspace
\Delta v_{\mu}=\frac{1}{\mu} \partial_{\mu} p , \enspace
 v^{\mu}(\vec{X}(t; \xi))= \dot{X}^{\mu}(t; \xi) ,
\end{equation}
where $p$ is the pressure from the fluid, and the last equation shows that there is no slipping between the surface of microorganism and the fluid.    

The action which gives these equations of motion is
\begin{eqnarray}
 \hat{S}_N & = & \sum_{i=1}^N \int dt \int d^{D-1}\xi_{(i)}
P_{\mu}^{(i)}(t; \xi_{(i)})
\left[ \dot{X}^{\mu}_{(i)}(t; \xi_{(i)}) - v^{\mu}(X_{(i)})
 \right] \nonumber \\
& &-{1 \over 2\pi\alpha'} \int d^D x \left[ {1 \over 4} \omega_{\mu\nu}
\omega^{\mu\nu} -{1 \over \mu} p(x) \partial_{\mu} v^{\mu}
\right] . \label{swimming}
\end{eqnarray}
where $\omega_{\mu \nu}=\partial_{\mu}v_{\nu}-\partial_{\nu}v_{\mu}$ 
denotes vorticity, and  $P^{(i)}_{\mu}$ is the Lagrange multiplier.  

From the viscous fluid, generation of the heat or the entropy occurs.  Knowing the rate of the entropy generation, we can estimate the statistical weight necessary to calculate the partition function.  Comparing the weight with the above action, the unknown constant $\alpha'$ is fixed;
\begin{equation}
{1 \over 4\pi\alpha'}={\mu t^{\ast} \over k_B T} .
\end{equation}
Here, $t^{\ast}$ is the period of the swimming motion.

This theory is identical to the Landau gauge QED, when the velocity vecter is considered as a gauge field.

Therefore, the swimming problem of the microorganisms has the closer relation to the $U(1)$ gauge theory coupled to the string or membrane variables, rather than to the hydrodynamics of Kalb-Ramond-Nambu given by
\begin{equation}
{\cal L}^{\star} = -\frac{1}{2}(v^{\mu})^2 - k W_{\mu\nu} \omega^{\mu\nu},
\end{equation}
where the velocity field is expressed in terms of the Kalb-Ramond field or the velocity potential as $v^{\mu} = \partial^{\mu}\tilde{W}_{\nu\mu}$.

The $U(1)$ gauge theory coupled to the string (membrane) is described by \begin{equation}
S^{\star} = \sum_{i=1}^{N} \int dt \int d\xi^{D-1}_{(i)} \dot{X}^{\mu}_{(i)} (t; \xi) v^{\mu}(X_{(i)}) - ( \mbox{  Landau gauge QED}) ,
\end{equation}
in which the interaction of the string (membrane) variable with the velocity field differs from that in the model of swimming of microorganisms in Eq.(\ref{swimming}). 

In the ordinary string and membrane theories, the velocity field $v^{\mu}$ couples to $\dot{X}^{\mu}(t; \xi)$, whereas in the microorganisms' swimming, $v^{\mu}$ couples to $P^{\mu}(t; \xi)$.  Since $\dot{X}^{\mu}$ and $P^{\mu}$ are Fourier conjugate with each other, which indicates that the string or membrane theory coupled to the $U(1)$ gauge field is, in a sence, dually related to the swimming of microorganisms.  

For example, the Casimir energy of the string (membrane) theory comes from the ultraviolet region, whereas in the swimming of microorganisms, it comes from the infrared region, and the most efficient swimming way is possible in the sum of the higher oscillation modes, which differs from the ordinary lower excitation modes of string or membrane.  This "duality" may also be understood from the difference of the boundary conditions: The no slipping condition for the microorganisms is the Dirichlet-type fixed boundary condition rather than the Neumann-type free boundary condition in the ordinary string or membrane theory.   
Please refer to the papers in Ref.~\cite{ours} for further details.  The references include also the relationship between the collective motion of $N-1$ microorganisms and the $N$ point Reggeon amplitude of string or membrane theory,  the generation of the red-tide and the liquid-gas phase transition, application of the area or volume preserving diffeomorphisms ($w_{1+\infty}$ and  $W_{1+\infty}$ algebras) to the swimming motion of microorganisms, and the swimming efficiency of the flagella and the cilia. 

A goal in this theme should be the understanding of why we have only three universal classes in the swimming ways of microorganisms; the swimming ways with the cilia, with the flagella, and with the bacterial flagella. 

%%%%%%%%%%%%%%%%%%%
\section{Conclusion}
(1) Old-fashioned dualities may be useful?
(2) Before concluding the usefulness of the old-fashioned dualities, the newly-fashioned dualities should be well studied.

\section*{Acknoledgements}
The author gives his sincere gratitude to the organizers of the Kashikojima Institute, especially Nobuhiro Maekawa, Taichiro Kugo and Masako Bando for giving me the opportunity of doing this lecture in enjoyable circumstances.  He deeply thanks Professor Seitaro Nakamura for inviting him to the Amagi-Highland Seminars, where the first trial of this lecture was given.  He is also grateful to Makoto Sakamoto for inviting him to the fruitful Yukawa Institute Workshop.  

The first part of this lecture is based on the author's works some 15 to 20 years ago, performed in collaboration with Koichi Seo and Masanori Okawa.  He deeply thanks them for their enthusiastic and reliable collaboration given on the dualities at that time.

The latter part is based on the works carried out in the past 5 years, he really thanks his collaborators on these works, Shin'ichi Nojiri, Masako Kawamura, Miyuki Katsuki, Hiroto Kubotani, Sergei Odintsov, and Emil Elizalde.  The earlier collaborations and discussions with Ichiro Oda, Akika Nakamichi and Fujie Nagamori on the topological field theory and gravity were very fruitful, from which the later works came out.

%%%%%%%%%%%%%%%%%%%%
%%%%%%%%%%%%%%%%%%

%%%%%%%%%%%%%%%%%%%%%%

\begin{thebibliography}{25}
\bibitem{Summer School} A. Sugamoto, The lecture note edited by the organizers, the then youth of the Kyoto U., of the 33th Summer School for Youth ('87) (in Japanese).
\end{thebibliography}
\end{document}